\documentclass[conference]{IEEEtran}

\IEEEoverridecommandlockouts

\usepackage[compress]{cite}

\mathchardef\citemidpenalty=1
\usepackage{microtype}
\usepackage{array}

\usepackage[breaklinks]{hyperref}

\usepackage[utf8]{inputenc}

\usepackage{url}

\usepackage{csquotes}

\usepackage{graphicx}
\graphicspath{{images/}}

\usepackage{multirow}

\usepackage{makecell}

\usepackage{array}

\usepackage{enumitem}

\usepackage{pifont}

\usepackage{tcolorbox}

\usepackage{listings}
\usepackage{balance}

\usepackage[labeled, resetlabels]{multibib}
\newcites{P}{Primary Studies}
\newcites{trash}{foo} 

\usepackage[compact]{titlesec}
\titlespacing*{\section}{0pt}{5pt}{4pt}
\titlespacing*{\subsection}{0pt}{3pt}{3pt}
\titlespacing*{\subsubsection}{0pt}{2pt}{2pt}

\AtBeginDocument{%
  \providecommand\BibTeX{{%
    \normalfont B\kern-0.5em{\scshape i\kern-0.25em b}\kern-0.8em\TeX}}}

\begin{document}

\title{Characterizing Technical Debt and Antipatterns in AI-Based Systems: A Systematic Mapping Study
{}\thanks{\IEEEauthorrefmark{1}All authors contributed equally to this study.}
}


\author{
    \IEEEauthorblockN{
        Justus Bogner\IEEEauthorrefmark{1}
	}
	\IEEEauthorblockA{
        University of Stuttgart\\
        Institute of Software Engineering\\
        Stuttgart, Germany\\
    	justus.bogner@iste.uni-stuttgart.de
    }
    \and
    \IEEEauthorblockN{
		Roberto Verdecchia\IEEEauthorrefmark{1}
	}
    \IEEEauthorblockA{
        Vrije Universiteit Amsterdam\\
        Department of Computer Science\\
        Amsterdam, The Netherlands\\
    	r.verdecchia@vu.nl
    }
    \and
    \IEEEauthorblockN{
		Ilias Gerostathopoulos\IEEEauthorrefmark{1}
	}
	\IEEEauthorblockA{
        Vrije Universiteit Amsterdam\\
        Department of Computer Science\\
        Amsterdam, The Netherlands\\
    	i.g.gerostathopoulos@vu.nl
    }
}


\thispagestyle{plain}
\pagestyle{plain}


\maketitle

\begin{abstract}
\textit{Background:}
With the rising popularity of Artificial Intelligence~(AI), there is a growing need to build large and complex AI-based systems in a cost-effective and manageable way.
Like with traditional software, Technical Debt~(TD) will emerge naturally over time in these systems, therefore leading to challenges and risks if not managed appropriately.
The influence of data science and the stochastic nature of AI-based systems may also lead to new types of TD or antipatterns, which are not yet fully understood by researchers and practitioners.
\textit{Objective:}
The goal of our study is to provide a clear overview and characterization of the types of TD (both established and new ones) that appear in AI-based systems, as well as the antipatterns and related solutions that have been proposed. 
\textit{Method:}
Following the process of a systematic mapping study, 21 primary studies are identified and analyzed. 
\textit{Results:}
Our results show that (i) established TD types, variations of them, and four new TD types (data, model, configuration, and ethics debt) are present in AI-based systems, (ii) 72 antipatterns are discussed in the literature, the majority related to data and model deficiencies, and (iii) 46 solutions have been proposed, either to address specific TD types, antipatterns, or TD in general.
\textit{Conclusions:}
Our results can support AI professionals with reasoning about and communicating aspects of TD present in their systems.
Additionally, they can serve as a foundation for future research to further our understanding of TD in AI-based~systems.
\end{abstract}

\begin{IEEEkeywords}
Artificial Intelligence, Machine Learning, Technical Debt, Antipatterns, Systematic Mapping Study
\end{IEEEkeywords}

\section{Introduction}

Artificial Intelligence (AI) covers different technologies for searching, reasoning, planning, problem solving, and learning with the overall aim of \enquote{automating intellectual tasks normally performed by humans}~\cite{francois_chollet_deep_2017}. 
Its rise in popularity in recent years is mostly due to advancements in Machine Learning (ML), an area of AI focusing on algorithms and systems to identify rules and patterns in data based on statistical modeling techniques.

With more and more companies offering AI-powered products and using AI techniques to improve their internal processes, there is a need to build large, complex AI systems in a cost-effective and manageable way. 
At the surface level, this may not seem like a new problem: AI systems are software systems too, so we can use well-known, established software engineering principles, practices, and processes to build such systems (e.g., separation of concerns, component-based encapsulation, and agile delivery). 
However, recent studies show that the AI/ML domain possesses characteristics that make it distinct from other software application domains, such as an increased importance of data quality and management, unclear abstraction boundaries for complex models, and challenges in the customization and reuse of AI/ML components~\cite{amershi_software_2019,rahman_machine_2019,Sculley2015}. 
Such characteristics seem to necessitate adaptations of principles, practices, and processes successfully used in other domains, or even the adoption of new, AI-specific ones~\cite{bosch_engineering_2021}.

A successful practice when building software systems in an iterative fashion is the awareness and management of technical debt~(TD)~\cite{RN2274}. 
TD is a metaphor used to describe design or implementation constructs that may be expedient in the short term, but can make future changes more costly or even impossible~\cite{avgeriou2016managing}.
Looking at the differences between AI/ML and other application domains from the TD perspective, researchers from Google proposed in 2015 new TD instances that are specific to the development of AI-based systems~\cite{Sculley2015}. 
Since this seminal paper, various research works from both academia and industry followed up with the documentation of additional TD items and antipatterns in AI/ML systems~\cite{agarwal2016making, breck2017ml,liu2020using}.

Despite these efforts, there is still no comprehensive conceptual overview of TD in AI-based systems. 
It is unclear, for instance, whether these systems accrue more \enquote{traditional} TD than other types of systems, such as code, architecture, or documentation debt. 
It is also important to understand if AI-specific TD types emerge and what their characteristics, associated antipatterns, and proposed solutions are.
Gaining such an overview would provide a foundation for future research, and support practitioners to better manage the maintenance and evolution of AI-based systems.
To characterize TD in AI-based systems, we therefore conducted a systematic mapping study (SMS), and collected and analyzed relevant papers on the topic.
Grounded in 21 primary studies, our contribution with this paper is the thorough analysis and discussion of the concepts of TD and antipatterns in AI-based systems.


\section{Background}
\label{sec:background}

\subsection{Technical Debt}
Since its formulation by Cunningham in 1992~\cite{RN2274}, the definition of \textit{technical debt} has continuously evolved and broadened in scope.
Nowadays, it encompasses a vast range of concepts, artifacts, and processes~\cite{fairbanks2020ur}.
Among the current definitions of TD, a widely adopted one was formulated during Dagstuhl seminar 16162~\cite{avgeriou2016managing}.
Simply referred to as the 16162 deﬁnition, it specifies TD as \enquote{design or implementation constructs that are expedient in the short term, but set up a technical context that can make a future change more costly or impossible}.
To structure the knowledge on TD, different \textit{TD types} have been described, e.g. architectural debt, requirements debt, and test debt, allowing researchers and practitioners to effectively focus on specific technical issues where the debt metaphor applies~\cite{li2015systematic}.
Instances of such types vary considerably in nature, from suboptimal reuse of architectural components~\cite{verdecchia2020architectural}, to deferred testing~\cite{guo2011portfolio}, or the involvement of certain development communities~\cite{tamburri2013social}.
In this research, we embrace the 16162 deﬁnition of TD, and build upon the TD type taxonomies presented by Li et al.~\cite{li2015systematic} and Rios et al.~\cite{rios2018tertiary}.

\subsection{Antipatterns}
While design patterns constitute proven solution blueprints for specific problems, there is also the inverse concept of \textit{antipatterns}, i.e. frequently occurring suboptimal solutions~\cite{Brown1998}.
Developers may choose these solutions under time pressure, but antipatterns often appear due to insufficient expertise.
They can have immediate negative effects on quality attributes such as maintainability, performance efficiency, or reliability, but may also hinder the sustainable evolution of a system, leading to the accumulation of TD.
Antipatterns can exist at different levels of abstraction, such as code antipatterns, architectural antipatterns, or even project management antipatterns.

There is also the related concept of \textit{bad smells}, e.g. code smells~\cite{Fowler1999} or architectural smells~\cite{Garcia2009}.
Some authors keep these terms strictly separated~\cite{Khomh2009}, i.e. software smells are seen as potential indicators of bad quality that may require further investigation, while an antipattern is always supposed to be a bad practice that should be avoided.
However, similar to patterns, many antipatterns can also be context-sensitive and may be perceived as \enquote{bad} only in specific cases.
When collecting archetypes of suboptimal software practices, a clear distinction between the two concepts becomes less important and several studies have handled them uniformly~\cite{Fontana2016,Bogner2019-CSEQUDOS}.
For the purpose of this paper, we therefore do not differentiate between the terms \textit{antipattern} and \textit{smell}, i.e. we collect both concepts under the same umbrella.
In this sense, we treat e.g. code smells as antipatterns on the implementation~level.

\subsection{Related Work}
Technical debt and antipatterns have been the target of numerous reviews in different SE subfields and domains.
However, to the best of our knowledge, there exists no comprehensive secondary study focusing on these concepts in the area of AI-based systems.
TD has been studied in the context of databases~\cite{Albarak2018} and data-intensive systems~\cite{Foidl2019a}, but without a clear focus on AI or ML, as with Sculley et al.~\cite{Sculley2015}.

Nonetheless, several studies focus on general software quality aspects in AI-based systems.
Humbatova et al.~\cite{Humbatova2020} conceptualized a fault taxonomy for deep learning systems by analyzing GitHub repositories, StackOverflow posts, and conducting interviews.
While faults are not in the scope of our study, they can be related to TD and antipatterns in some cases, e.g. as symptoms of their existence.
Concerning the quality assurance of AI-based systems, several position papers discuss differences compared to \enquote{traditional} systems and highlight the need for adapted quality assurance techniques~\cite{Felderer2021,Santhanam2020}.
Some empirical studies also went further than this and distilled effective techniques for ML system engineering, e.g. Serban et al.~\cite{Serban2020} derived general AI system development best practices, and Siebert et al.~\cite{Siebert2020} provided guidelines for the quality assurance of such systems.
Lastly, the study that comes closest to the goal of our own research is a preliminary multivocal literature review by Washizaki et al.~\cite{Washizaki2019}:
they collected design patterns and antipatterns for ML systems from both white and grey literature.
However, they only identified eight antipatterns, seven from Sculley et al.~\cite{Sculley2015} and one from a company blog post, and did not provide any insights on how TD is characterized in AI-based systems.
With our study, we therefore fill this gap by providing a detailed characterization of TD and antipatterns in AI-based systems.

\begin{figure*}
    \centering
    \includegraphics[width=1\textwidth]{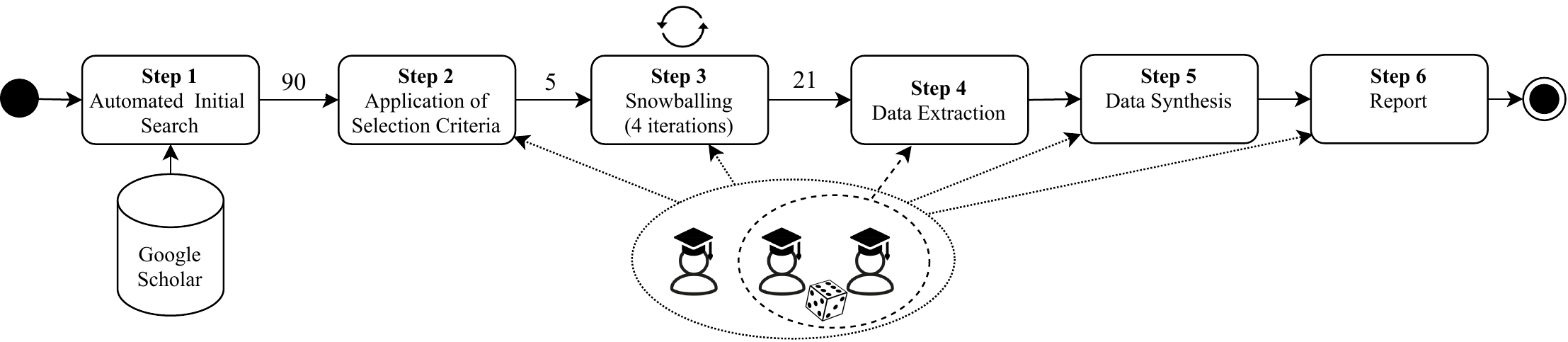}
    \caption{Systematic mapping study process overview}
    \label{fig:process}
\end{figure*}

\section{Methodology}
\label{sec:method}
In this section, we document the research design, which was rigorously adhered to during study execution.
We primarily followed the guidelines for conducting systematic literature studies in software engineering research by Kitchenham~\cite{Kitchenham2004}.

\subsection{Research Objective and Questions}
\label{sec:objAndRqs}
The aim of this research is to further the understanding of technical debt and antipatterns in AI-based software systems.
To refine this goal, we derived the following research questions (RQs), which guided our mapping study:

\begin{enumerate}[label=\textbf{RQ\arabic*:},leftmargin=*]
    \item What are the characteristics of technical debt in AI-based systems?
    \begin{enumerate}[label=\textbf{RQ1.\arabic*:},leftmargin=10pt]
        \item Which established types of technical debt have been reported for AI-based systems?
        \item Does the nature of established technical debt types change in AI-based systems?
        \item Which new technical debt types have emerged in AI-based systems?
        \item Which quality attributes are affected by technical debt in AI-based systems?
    \end{enumerate}
    \item Which antipatterns have been reported for AI-based systems?
    \item Which solutions have been reported to address technical debt and antipatterns in AI-based systems?
\end{enumerate}


\subsection{Research Process}
An overview of the research process followed is depicted in Fig.~\ref{fig:process}.
The process started with the execution of a conservative automated search query via \textit{Google Scholar}, followed by an iterative forward- and backward-snowballing process, until theoretical saturation was reached.
Following the methodology by Wohlin et al.~\cite{Wohlin2014}, we based our search on a start set obtained via an automated search query executed on Google Scholar.
This set was then used for exhaustive bidirectional snowballing.
The use of Google Scholar allowed us to avoid bias in favor of any specific publisher~\cite{Wohlin2014}.
Further details of each research step are reported in the following subsections.

\subsubsection{Step 1: Automated Initial Search}
To gather an initial set of potentially relevant studies, we executed a conservative automated query on the \textit{Google Scholar} literature indexer.
The title-focused query was designed to encompass related literature focusing specifically on the topic under investigation, and is formulated as follows:

\begin{center}
\vspace{-8pt}
\begin{tabular}{c}
\begin{lstlisting}[language=sql, columns=fullflexible, caption={Automated search query}, numbers=left, label=list:query]
(*\bfseries ALLINTITLE*): ("technical debt" OR "antipatterns" OR 
"antipattern" OR "anti patterns" OR "anti pattern" OR 
"smell" OR "smells") AND ("artificial intelligence" OR 
"machine learning" OR "deep(*\textvisiblespace*)learning" OR "intelligent" 
OR "smart" OR "AI" OR "ML" OR "DL")
\end{lstlisting}
\end{tabular}
\end{center}

This query identifies literature containing in their title keywords referring either to \enquote{technical debt}, \enquote{antipatterns}, or \enquote{smells} (Listing~\ref{list:query}, Lines 1-3) and keywords referring to AI, or related synonyms and acronyms (Listing~\ref{list:query}, Lines 3-5).
The automated query was executed mid-June 2020, and yielded 90 potentially relevant studies. As we were not interested in publications regarding a specific timeframe, the publication date was purposely left unbounded in the query.

\subsubsection{Step 2: Application of Selection Criteria}
After identifying the set of potentially relevant studies via the automated query, we conducted a manual selection process. During this step, we evaluated the initial pool of studies based on pre-defined selection criteria.
A paper was selected as a primary study if it satisfied all inclusion criteria and none of the exclusion ones.
We used the following criteria:

\begin{itemize}
    \item[I1-] Publications reporting technical debt, antipatterns, or suboptimal software engineering practices
    \item[I2-] Publications focusing on AI-based systems
    \item[E1-] Non-English publications
    \item[E2-] Publications for which the full text is not available to us    
    \item[E3-] Duplicates or extensions of already included publications
    \item[E4-] Secondary or tertiary studies
    \item[E5-] Publications in the form of editorials, tutorials, books, extended abstracts, etc.
    \item[E6-] Non-scientific publications (i.e. grey literature)
\end{itemize}

The two inclusion criteria (I1, I2) were formulated to ensure that primary studies focused on the investigated topic, namely TD and antipatterns in AI-based systems, and hence provided relevant data to answer our RQs.
The exclusion criteria instead were designed to guarantee that data could be extracted from papers (E1, E2), without duplication or redundancy (E3, E4), and consisted of scientific literature (E5,~E6).

Given the fast pace at which the investigated topic evolves, we purposely included preprints during the selection process.
However, preprints needed to possess a sufficient level of quality (reviewed by all three researchers), to have already been cited by high-quality academic literature, and to be from reputable authors who published other studies in the field.\footnote{This design decision led to the inclusion of two additional papers, namely a white paper by Microsoft research~\citeP{agarwal2016makingprimary}, and a paper presented at the \textit{AAAI Fall Symposium Series: Artificial Intelligence in Government and Public Sector}~\citeP{lewis2019component}.}

During the selection, adaptive reading depth~\cite{Petersen2008} was used to efficiently assess potentially relevant studies.
To mitigate subjective bias, all authors independently applied the selection criteria for the 90 candidate studies.
Differences were jointly discussed until consensus was reached.
This led to the selection of five primary studies, i.e. the snowballing start set.

\subsubsection{Step 3: Snowballing}
To obtain a sound and encompassing set of primary studies, the automated search was complemented by recursive backward and forward snowballing~\cite{Wohlin2014}.
During this step, all studies either citing or cited by the primary studies were examined.
Similar as for the initial selection, the snowballing process was conducted by three researchers: in each round, all researchers independently suggested new primary studies to be included, i.e. studies which fulfilled the selection criteria.
Divergences were jointly discussed and resolved, after which the next iteration started.
Overall, it took four rounds of backward and forward snowballing until no new studies were identified.
Snowballing led to the inclusion of 16 studies, i.e. our SMS selection process led to the identification of 21 primary studies.

\subsubsection{Step 4: Data Extraction}
\label{sec:data_extraction}
In the next step, we systematically analyzed the primary studies and extracted data related to our RQs.
To gain a preliminary understanding, a data extraction pilot with four papers was conducted independently by all researchers.
Subsequently, the extracted data was jointly discussed, leading to the extraction framework used in this study.
Two researchers were randomly assigned to each primary study.
They independently extracted the data and agreed on the final extractions per paper in a consensus meeting, with the intervention of the third researcher when required.
The extraction framework is divided into one part for each specific RQ of our study (see Section~\ref{sec:objAndRqs}).

To \textit{characterize TD in AI-based systems} (RQ1), data needed to be extracted according to its four sub-questions.
For the recurrence of \textit{established TD types in AI-based systems} (RQ1.1), we identified types based on the TD taxonomies of Li et al.~\cite{li2015systematic} and Rios et al.~\cite{rios2018tertiary} (e.g. code, test, or architectural debt).
Using the same taxonomies, we also extracted \textit{variations of established TD types} (RQ1.2), i.e.
if the nature or scope of TD types changed in AI-based systems.
An example for this is test debt, as it extends in AI-based systems to testing models and data.
We also analyzed the primary studies for \textit{new TD types in AI-based systems} (RQ1.3), i.e. debt types which (i) are documented in the context of AI-based systems and (ii) cannot be traced back to established~TD~types.
Lastly, we extracted \textit{quality attributes affected by TD in these systems} (RQ1.4), which was based on ISO/IEC 25010~\cite{ISO25010}, with the possibility to extend it with additional identified attributes.

To answer RQ2 (\textit{antipatterns}), the primary studies were analyzed for recurrent suboptimal solutions in AI-based systems.
Such suboptimal solutions could be explicitly referenced as \enquote{antipatterns} (e.g. correction cascades~\cite{Sculley2015}), or reported as root causes of TD (e.g. unstable data dependencies~\cite{liu2020using}). 

Finally, to answer RQ3 (\textit{solutions}), we extracted the solutions proposed to mitigate or resolve TD and antipatterns in AI-based systems.
Such solutions could be specific to a certain TD type or antipattern (e.g. model isolation to resolve entanglements) or general best practices to mitigate or prevent the introduction of TD in AI-based systems (e.g. periodically assessing assumptions during ML model evolution).

Note that the extractions did not have to explicitly mention \enquote{debt}, \enquote{antipattern}, or a specific quality attribute.
The decision if a passage implicitly describing these concepts warranted extraction was up to the researchers' interpretation.

\subsubsection{Step 5: Data Synthesis}
As a final step, the extracted data was harmonized (e.g. merging identical or very similar antipatterns), and then analyzed to derive answers to the research questions.
This analysis relied on open coding~\cite{jenner2004companion} to systematically identify recurrent concepts.
Further axial coding~\cite{jenner2004companion} was required to reduce the growing complexity of some emerging concepts (e.g. antipattern subcategories).
During the coding, emerging results were continuously discussed among the authors to keep codes and their abstraction level consistent.
Finally, summary statistics were created to discuss general findings and their potential implications.
For the sake of transparency and reproducibility, we make all study artifacts publicly available online\footnote{\url{https://doi.org/10.5281/zenodo.4457216}}.

\section{Results}
\label{sec:results}
Our results are extracted from 21 primary studies \citeP{agarwal2016makingprimary}-[P21], which were published in conferences (12/21), workshops (5/21), journals (3/21), or distributed as white papers (1/21). 
Since the appearance of the first paper focusing on TD in AI-based systems in 2015~\citeP{Sculley2015primary}, we observed a growing publication trend until 2020\footnote{Year in which the primary study selection was executed.}. 
Interestingly, a large number of primary studies were co-authored by at least one industrial practitioner (13/21), including nine papers authored exclusively by practitioners. 
Google is the most recurrent company~\citeP{Sculley2015primary, breck2017primary, o2020common, polyzotis2019data, hynes2017data}, while other prominent examples include Microsoft~\citeP{agarwal2016makingprimary}, Amazon~\citeP{schelter2018challenges}, and IBM~\citeP{shrivastava2019dqa}. 
The considerable involvement of industrial parties displays the industrial relevance of the topic, which has still to gain traction in academic environments. 

In the remainder of this section, we present the results of our study, according to the four RQs guiding the investigation.

\subsection{Characteristics of TD in AI-based Systems (RQ1)}
\label{sec:rq1}
\begin{figure}[hbpt!]
    \centering
    \hspace{-5pt}
    \includegraphics[width=0.5\textwidth]{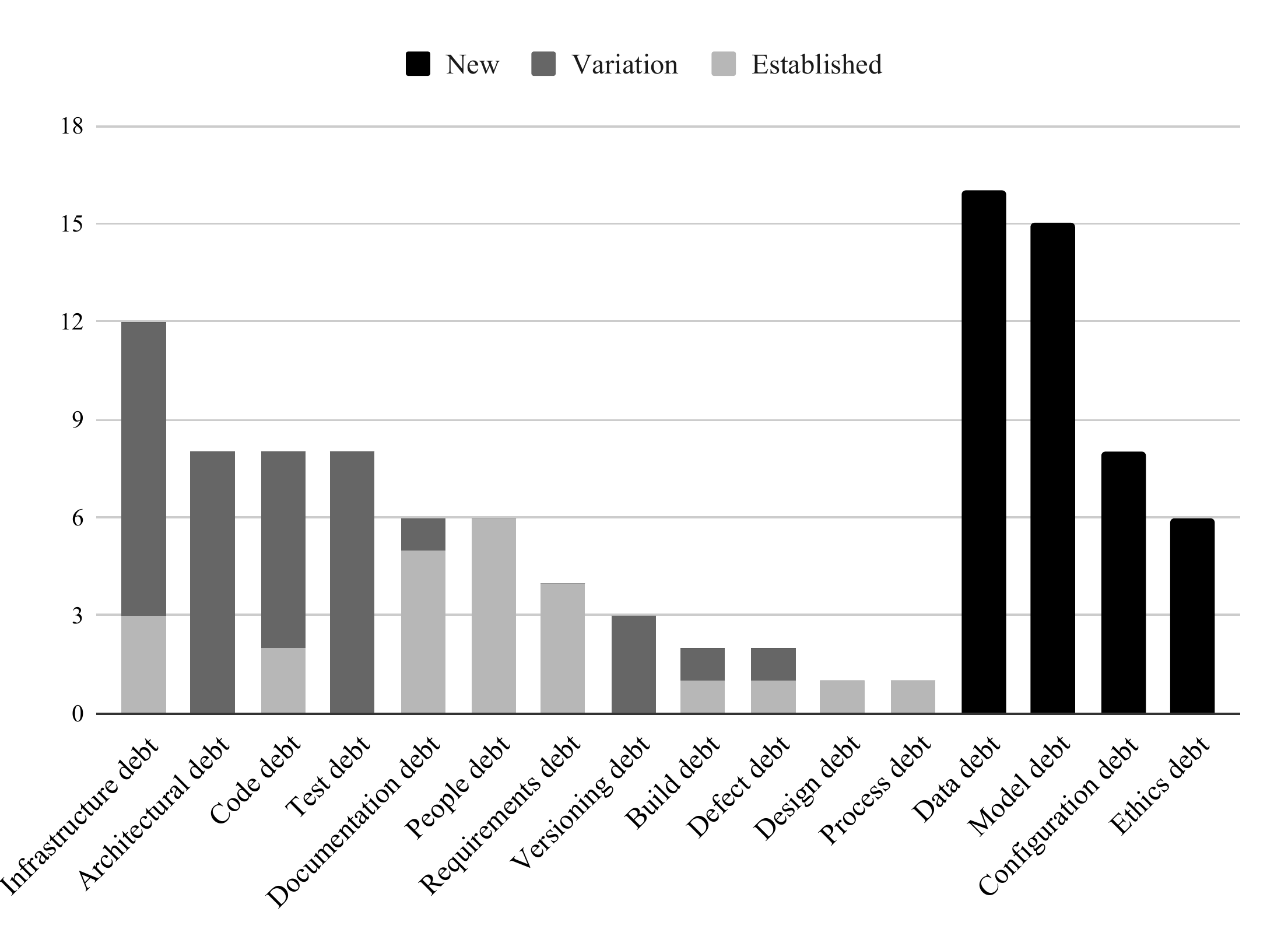}
    \caption{Recurrence of TD types in AI-based systems\vspace{-15pt}}
    \label{fig:tdtypes}
\end{figure}

This section reports the results for the sub-questions of RQ1, aiming to characterize the nature of TD in AI-based systems.
An overview of the recurrence of all identified TD types is reported in Fig.~\ref{fig:tdtypes} (both established and new types).
Following we discuss the distribution of established TD types plus their variations, new types emerging in AI-based systems, and finally affected quality attributes.
Mentioned antipatterns are explained in more detail in Section~\ref{sec:antipatterns}.

\subsubsection{Established TD Types and Variations (RQ1.1 \& RQ1.2)}
As shown in Fig.~\ref{fig:tdtypes}, infrastructure debt is the most recurrent established TD type (12/21), followed by architectural, code, and test debt (8/21).
Other types of debt, such as documentation~(6/21), people~(6/21), requirements~(4/21), and versioning debt~(3/21) are also reported, albeit less frequently.
Lastly, build (2/21), defect (2/21), design (1/21), and process debt (1/21) are only sporadically mentioned.

We observed that this distribution is mostly due to new engineering challenges related to AI-based systems.
Our primary studies link these systems to an inherently experimental development process, deep entanglement between architectural components and utilized data, and necessary data transformation steps between components.
Coping with such difficulties often leads to the introduction of TD, e.g. the evolutionary composition of AI pipelines may result in general-purpose components being precariously stitched together via \textit{glue code}~\citeP{Sculley2015primary}.
Additionally, the data-driven nature of AI algorithms introduces novel difficulties for quality assurance, often manifested as suboptimal testing practices, such as under-tested or ill-tested functionalities.

Moreover, AI-specific variations or extensions of TD types, rather than their established scope~\cite{li2015systematic, rios2018tertiary}, are often used in our primary studies.
Specifically, out of the 61 established TD type extractions, 37 are such variations.
While some TD types occur unchanged in AI-based systems, re-interpreting several existing debt types is necessary to accommodate the new characteristics of these systems.
The most recurrent established debt types, namely infrastructure, architectural, code, and test debt, are also the types which frequently exhibit an extended or augmented scope in our primary studies.

Regarding \textit{infrastructure debt}, this TD type is extended with deficiencies related to the implementation and operation of AI pipelines as well as to the management of AI models.
In AI-based systems, this TD type often manifests in form of complex infrastructure comprising various AI pipelines~\citeP{Sculley2015primary}, suboptimal allocation of resources to train/test AI models~\citeP{lewis2019component}, and weak AI monitoring and debugging capabilities~\citeP{agarwal2016makingprimary}, leading to major operations and reproducibility~issues.

\textit{Architectural debt} variations instead reflect the emphasis in AI-based systems on data, leading to a deep entanglement of architecture components with their underlying data.
This may introduce debt items such as complex and non-deterministic dependencies between architectural components and datasets~\citeP{lewis2019component}, hard to assess compositions of architectural elements~\citeP{liu2020primary}, or \textit{undeclared consumers}~of~AI~models~\citeP{belani2019requirements}.

Similarly, \textit{test debt} extends to data testing, ranging from naive omission of basic sanity checks to the lack of more sophisticated tests to assess data quality or distributions~\citeP{Sculley2015primary}.
In addition, new facets include suboptimal practices in testing AI models and pipelines.
Their deep connection to training data and the stochastic nature of some AI algorithms~\citeP{polyzotis2019data, breck2017primary} make these artifacts increasingly complex to evaluate.

\textit{Code debt} is shaped by the experimental nature of AI model development.
This frequently emerges in form of \textit{dead experimental code paths} in production code~\citeP{arpteg2018software} and the suboptimal refactoring of experimental models into deployable software~\citeP{o2020common}.
The algorithmic complexity of these systems also increases the likelihood of certain code deficiencies~\citeP{jebnoun2020scent}.

For \textit{versioning debt} as a less referenced TD type, we exclusively identified the AI-centric usage of the term.
In AI-based systems, this now includes the versioning of AI models and training/testing data, which is often done in suboptimal fashion, if at all~\citeP{arpteg2018software, schelter2018challenges, breck2017primary}.

Finally, documentation, people, requirements, build, defect, design, and process debt mostly appear in our studies according to their established scope, i.e. while such debt types also appear in AI-based systems, characteristics of AI do not have a prominent impact on their manifestation.
This highlights that numerous commonalities are shared with software systems not employing AI.
Missing documentation, insufficient developer skills, and unclear system requirements are all examples of TD which also frequently occur in non-AI software.
Slight variations for some of these TD types are the extension of documentation debt to features and assumptions on the used data~\citeP{matthews2020patterns}, of build debt to suboptimal dependencies of internal and external AI models~\citeP{liu2020primary}, and of defect debt to ignored issues related to the quality of model predictions~\citeP{breck2017primary}.

\subsubsection{New TD types in AI-based systems (RQ1.3)}
With this RQ, we wanted to synthesize new TD types important for AI-based systems, i.e. types not included in the taxonomies of Li et al.~\cite{li2015systematic} and Rios et al.~\cite{rios2018tertiary}.
Specifically, we found four such types of debt: \textit{data debt}, \textit{model debt}, \textit{configuration debt}, and \textit{ethics debt}, which we further~describe~below.

\textit{Data debt.}
The most recurrent new TD type regards suboptimal constructs around the data used in AI-based systems (16/21).
Specifically, this TD type refers to deficiencies related to the collection, management, and usage of data, both for training and production~\citeP{munappy2019data,hynes2017data,gudivada2017data}.
In addition to causing immediate issues, this TD type can also be latent, i.e. not manifesting itself immediately, but rather posing a risks for the long-term evolution of systems.
Commonly referenced instances of data debt are data quality issues, unmanaged data dependencies and anomalies, or poor data relevance.
Given their heavy reliance on data, this TD type can strongly impact AI systems, including reduced classification effectiveness, data loss due to \textit{premature aggregation}~\citeP{munappy2019data}, and compatibility~issues.

\textit{Model debt.}
The second most referenced new TD type is model debt (15/21).
This AI-specific debt type regards suboptimal practices in the design, training, and management of AI models~\citeP{breck2017primary,schelter2018challenges,Sculley2015primary}.
As such, model debt manifests itself as deficiencies occurring exclusively in model-related constructs. Most prominently, model debt originates from suboptimal feature selection processes, neglected hyperparameter tuning, and poorly engineered model deployment strategies.  
Recurrent items of model debt are \textit{feature entanglement}~\citeP{Sculley2015primary}, \textit{hidden feedback loops}~\citeP{alahdab2019empirical}, \textit{unrecognized model staleness}~\citeP{schelter2018challenges}, and substantial differences between training and production performance, i.e. \textit{training/serving skew}~\citeP{Sculley2015primary}.
As models constitute the logic kernel of AI-based systems, this TD type can have serious consequences, ranging from major challenges in maintaining a model, to severe deterioration of model~accuracy.

\textit{Configuration debt.}
This debt type (8/21) describes deficiencies around the configuration mechanisms of AI-based systems~\citeP{alahdab2019empirical,arpteg2018software,Sculley2015primary}.
Often, configuration debt arises when the complexity of e.g. dynamic feature selection, hyperparameter tuning, and data pre- and post-processing makes it difficult to efficiently outsource machine- and human-readable configuration files for these activities.
This encourages AI engineers to take shortcuts and to only consider the clean-up, restructuring, and commenting of configuration files as an afterthought.
While the lines of configuration for AI-based systems may even exceed the lines of source code~\citeP{Sculley2015primary}, configurations are frequently not given the same level of quality control as code, e.g. reviews or tests.
Prominent instances of configuration debt include massive/complex, poorly documented (or simply undocumented), unversioned, or untested configuration files.
As such, configuration debt may have some touch points with e.g. infrastructure or documentation debt, but its explicit description in the context of AI still warrants its own TD type.

\textit{Ethics debt.}
One less referenced new debt type is ethics debt (6/21), which comprises deficiencies around ethical aspects of AI-based systems, such as algorithmic fairness, prediction bias, or a lack of transparency and accountability~\citeP{vakkuri2020just,matthews2020patterns,roselli2019managing}.
Specifically, ethics debt arises when socio-ethical concerns are deliberately or inadvertently neglected during the design or training phase of AI-based systems.
While this can go hand in hand with reduced model accuracy, the resulting systems may also satisfy all technical requirements while leaving one or more ethical concerns unaddressed.
This debt type can lead to ethical fallacies so deeply embedded into an AI-based system that they usually cannot be resolved with only minor data or software changes.
Instead, they may require a complete restructuring and retraining of AI models or major source code updates.
Depending on the relevant regulations, ethics debt can also have legal consequences.

\subsubsection{Affected quality attributes (RQ1.4)}
\begin{figure}[hbpt!]
    \centering
    \hspace{-5pt}
    \includegraphics[width=0.5\textwidth]{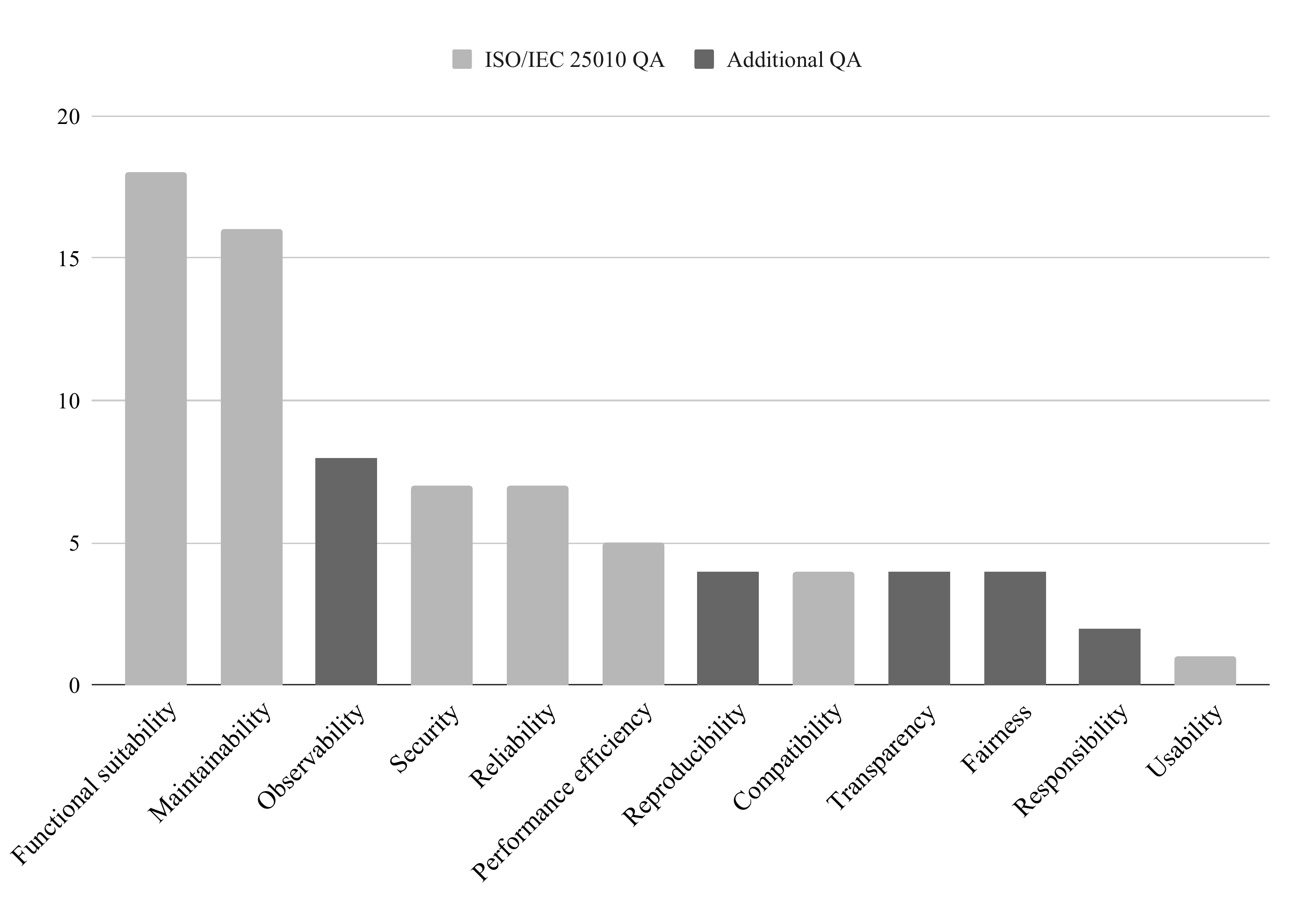}
    \vspace{-20pt}
    \caption{Recurrence of quality attributes impacted by TD in AI-based systems\vspace{-12pt}}
    \label{fig:qa}
\end{figure}

We extracted a total of 12 unique quality attributes impacted by TD in AI-based systems, for which the recurrence in our 21 primary studies is depicted in Fig.~\ref{fig:qa}.

Overall, functional suitability is the most referenced quality attribute (18/21).
Specifically, the introduction of TD in AI-based systems leads to issues in the functional correctness sub-characteristic (18/21), which, given the data-driven nature of AI-based systems, can be cumbersome to detect and resolve.
Among the most mentioned reasons for this is diminished model accuracy, e.g., due to \textit{training/serving skew}~\citeP{breck2017primary}.
Maintainability is the second most mentioned quality attribute (16/21), with emphasis on the sub-characteristics modifiability (14/21), testability (13/21), and reusability (11/21).
In this case, we noted how TD frequently leads to quality issues specific to or caused by AI models, such as difficult model re-training or reuse, ripple effects on changes through harmful dependencies, or complex collections of scripts and pipelines that are hard to analyze.
Observability (8/21) as additional quality attribute represents the degree to which the runtime behavior of a deployed AI-based system can be monitored.
This quality attribute can be severely impacted by infrastructure and architecture debt.
Security and reliability are referenced equally often (7/21).
Security issues in AI-based systems often concern the sub-characteristics accountability (5/21) or confidentiality (4/21).
Reliability, by contrast, is most often related to maturity of AI-based systems, i.e. the degree to which the system meets expectations under normal operation.
Other quality attributes, such as performance efficiency (5/21), compatibility (4/21), and usability (1/21), are mentioned less frequently, i.e. are most likely less impacted by TD in AI-based systems.
Similarly, new quality attributes emerging in AI-based systems such as reproducibility~(4/21), fairness (4/21), transparency~(4/21), and responsibility~(3/21) are also not frequent.
We believe the growing attention to topics related to these quality attributes in academia and industry will probably lead to a higher recurrence in the near future.
Finally, the only attribute from ISO/IEC 25010 not mentioned at all is portability.

\begin{tcolorbox}
\textbf{Main findings (RQ1):} Infrastructure debt (12) is the most recurrent established TD type in AI-based systems, followed by architectural, code, and test debt (8). We identified four new debt types emerging in AI-based systems, namely data, model, configuration, and ethics debt. Functional suitability (18) and maintainability (16) are the most impacted quality attributes, followed by observability (8), security (7), and reliability (7).
\end{tcolorbox}

\subsection{Antipatterns (RQ2)}
\label{sec:antipatterns}
From the 21 primary studies, we extracted a total of 72~unique antipatterns for AI-based systems.
To organize this large collection, we formed six categories (with an additional level of subcategories for the larger ones), where each antipattern is assigned to exactly one category.
These categories, and associated number of antipatterns, are displayed in Table~\ref{table:antipatternCategories}, while Table~\ref{table:antipatterns} lists the 14 most prominent antipatterns mentioned in at least three publications.
The largest categories are \textit{model antipatterns} (29) and \textit{data antipatterns} (22), which together account for $\sim$70\% of all identified antipatterns.
The four remaining categories -- \textit{design \& architecture antipatterns} (8), \textit{code antipatterns} (5), \textit{infrastructure antipatterns} (4), and \textit{ethics antipatterns} (4) -- are all of similar size.
In the following subsections, we briefly present each category with some antipattern examples.

\begin{table}
    \centering
	\caption{Antipattern categories, in parentheses: \# of unique antipatterns per (sub)category\vspace{-3pt}}
	\label{table:antipatternCategories}
	\begin{tabular}{p{0.2175\columnwidth}p{0.3265\columnwidth}p{0.296\columnwidth}}
		Category & Subcategory & Sources\\
		\hline
		\hline
		\multirow{5}{*}{Model (29)} & Training (9) &
		\makecell*[{{p{0.296\columnwidth}}}]{\hspace*{0.5pt}\citeP{Sculley2015primary,breck2017primary,matthews2020patterns,roselli2019managing}\newline\hspace*{0.5pt}\citeP{lwakatare2019taxonomy,munappy2019data,foidl2019risk}}\\
		 & Training/serving skew (6) &
		 \makecell*[{{p{0.296\columnwidth}}}]{\hspace*{0.5pt}\citeP{breck2017primary,matthews2020patterns,arpteg2018software,roselli2019managing}\newline~\hspace*{0.5pt}\citeP{lewis2019component,polyzotis2019data,lwakatare2019taxonomy,munappy2019data}}\\
		 & Features (6) & \citeP{Sculley2015primary,breck2017primary,alahdab2019empirical,roselli2019managing,lwakatare2019taxonomy,belani2019requirements}\\
		 & Management (4) & \citeP{breck2017primary,matthews2020patterns,schelter2018challenges}\\
		 & Validation (4) & \citeP{Sculley2015primary,breck2017primary,roselli2019managing,vakkuri2020just,lwakatare2019taxonomy}\\
		\hline
		\multirow{5}{*}{Data (22)} & Management (7) & \citeP{breck2017primary,arpteg2018software,o2020common,polyzotis2019data,munappy2019data,gudivada2017data}\\
		 & Anomalies (6) & \citeP{hynes2017data,foidl2019risk,shrivastava2019dqa}\\
		 & Quality (4) & \makecell*[{{p{0.296\columnwidth}}}]{\citeP{Sculley2015primary,matthews2020patterns,hynes2017data,munappy2019data}\newline\hspace*{0.2pt}\citeP{foidl2019risk,gudivada2017data,shrivastava2019dqa}}\\
		 & Relevance (3) & \citeP{matthews2020patterns,roselli2019managing}\\
		 & Dependencies (2) & \citeP{Sculley2015primary,liu2020primary,belani2019requirements}\\
		\hline
		\multirow{3}{*}{\makecell[l]{Design \&\\architecture (8)}} & Modularity (4) & \citeP{Sculley2015primary,matthews2020patterns,o2020common,lwakatare2019taxonomy,belani2019requirements}\\
		 & Integration (2) & \citeP{Sculley2015primary,alahdab2019empirical,lewis2019component}\\
		 & Technology adoption (2) & \citeP{Sculley2015primary,alahdab2019empirical,schelter2018challenges,o2020common}\\
		\hline
		\multirow{2}{*}{Code (5)} & Recurrent in AI (3) & \citeP{jebnoun2020scent}\\
		 & AI-specific (2) & \citeP{Sculley2015primary,alahdab2019empirical,arpteg2018software,roselli2019managing}\\
		\hline
		Infrastructure (4) & -- & \citeP{breck2017primary,arpteg2018software,agarwal2016makingprimary,lewis2019component,lwakatare2019taxonomy}\\
		\hline
		Ethics (4) & -- & \citeP{breck2017primary,matthews2020patterns,alahdab2019empirical,vakkuri2020just}\\
		\hline
	\end{tabular}
\end{table}

\begin{table}
	\caption{Most referenced antipatterns (occurred in at least 3 sources)\vspace{-3pt}}
	\label{table:antipatterns}
	\begin{tabular}{lll}
		Antipattern Name & Category & Sources\\
		\hline
		\hline
		Training/serving skew & Model & \makecell*[{{p{0.296\columnwidth}}}]{\hspace*{-2.5pt}\citeP{breck2017primary, arpteg2018software, roselli2019managing} \newline\citeP{lewis2019component, lwakatare2019taxonomy, munappy2019data}}\\
        Data duplication & Data & \citeP{hynes2017data, munappy2019data, gudivada2017data, shrivastava2019dqa}\\
        Data miscoding smell & Data & \citeP{Sculley2015primary, hynes2017data, munappy2019data, foidl2019risk}\\
        Null/missing data values & Data & \citeP{hynes2017data, munappy2019data, gudivada2017data, shrivastava2019dqa}\\
        Feature entanglement & Model & \citeP{Sculley2015primary, alahdab2019empirical, lwakatare2019taxonomy, belani2019requirements}\\
        Dead experimental codepaths & Code & \citeP{Sculley2015primary, alahdab2019empirical, arpteg2018software}\\
        Unstable data dependencies & Data & \citeP{Sculley2015primary, liu2020primary, belani2019requirements}\\
        Unsound/missing metadata & Data & \citeP{polyzotis2019data, munappy2019data, gudivada2017data}\\
        Glue code & \makecell[l]{Design \&\\architecture} & \citeP{Sculley2015primary, alahdab2019empirical, lewis2019component}\\
        Undeclared consumers & \makecell[l]{Design \&\\architecture} & \citeP{Sculley2015primary, lwakatare2019taxonomy, belani2019requirements}\\
        Multiple-language smell & \makecell[l]{Design \&\\architecture} & \citeP{Sculley2015primary, alahdab2019empirical, schelter2018challenges}\\
        Correction cascades & \makecell[l]{Design \&\\architecture} & \citeP{Sculley2015primary, matthews2020patterns, belani2019requirements}\\
        Weak or missing monitoring & Infrastructure & \citeP{breck2017primary, agarwal2016makingprimary, lewis2019component}\\
        Direct feedback loops & Model & \citeP{Sculley2015primary, lwakatare2019taxonomy, munappy2019data}\\
		\hline
	\end{tabular}
\end{table}

\textit{Model Antipatterns.}
The largest group of identified antipatterns (29) describes deficiencies or suboptimal practices with AI models, mostly in the context of machine learning.
Often, these refer to the training, validation, or management of models, i.e. to specific activities in the model life cycle.
A concrete example is the training antipattern \textit{direct feedback loops}~\citeP{Sculley2015primary, lwakatare2019taxonomy, munappy2019data}, which describes the unwanted state that a model impacts its own future training data selection.
This self-sustaining relationship may lead to wrong classifications and biased decisions, especially if feedback loops remain hidden or are not managed appropriately.
An example for a model validation antipattern is \textit{offline/online proxy metric divergence}~\citeP{breck2017primary, lwakatare2019taxonomy}.
Effectiveness of a production system is usually evaluated with metrics like user engagement or revenue (online), while models of AI components in such a system are validated with e.g. accuracy or mean squared error (offline).
If offline metrics are not sufficiently aligned with the online metrics, e.g. via correlation, system effectiveness may be severely impacted.
Lastly, a general model management antipattern is the absence of versioning and version control systems specifically for the model (\textit{no version control for models}~\citeP{breck2017primary}), which hinders sustainable evolution and potential rollbacks.

A different type of antipatterns in this category focuses on model features and their relationships.
An example of this is \textit{feature entanglement}~\citeP{Sculley2015primary, alahdab2019empirical, lwakatare2019taxonomy, belani2019requirements}, which refers to the interdependence of different model features.
Adding, removing, or changing the distribution of one feature often has an impact on other features, therefore hindering incremental system improvement.
This has also been described as the CACE principle~\citeP{Sculley2015primary}, i.e. \enquote{changing anything changes everything}.
A second example are \textit{epsilon features}~\citeP{breck2017primary, alahdab2019empirical}, which are features only leading to negligible model improvement.
Since every feature comes with costs for maintenance and evolution (especially when considering \textit{feature entanglement}), feature inclusion should be carefully considered based on merit.

Finally, a frequently mentioned type of model antipatterns is related to \textit{training/serving skew}~\citeP{breck2017primary, arpteg2018software, roselli2019managing, lewis2019component, lwakatare2019taxonomy, munappy2019data}.
This is the most recurrent unique antipattern and, in general, describes substantial model accuracy divergences in the production system when compared to the training accuracy.
Reasons for this can be different code paths to compute features in production, but also a non-representative or non-exhaustive training data set.
More specialized variants of this antipattern have been called \textit{distribution skew} or \textit{scoring/serving skew}~\citeP{polyzotis2019data}.
The divergence between training and serving accuracy can also emerge slowly over time, which is called \textit{data drift}~\citeP{lwakatare2019taxonomy, munappy2019data} and leads to \textit{stale models}~\citeP{breck2017primary,matthews2020patterns}.

\textit{Data Antipatterns.}
The second largest antipattern group ~(22) is related to deficiencies or suboptimal practices around the data of AI-based systems.
Since data is the foundation for machine learning models, such antipatterns can substantially diminish system effectiveness.
Many instances in this area are related to data quality or the existence of data anomalies.
Examples of bad quality are \textit{data duplication}~\citeP{hynes2017data, munappy2019data, gudivada2017data, shrivastava2019dqa}, \textit{null/missing data values}~\citeP{hynes2017data, munappy2019data, gudivada2017data, shrivastava2019dqa}, or \textit{data miscoding smells}~\citeP{Sculley2015primary, hynes2017data, munappy2019data, foidl2019risk}, where an attribute is represented with an unsuitable data type or format.
Similarly, examples for anomalies in machine learning data can be an \textit{unnormalized feature}~\citeP{hynes2017data, foidl2019risk}, where values exhibit large variance, or very few \textit{extreme outliers}~\citeP{hynes2017data}, which may distort important aggregate values used by models.
While of both these subcategories can also be important for data-intensive non-AI systems, these antipatterns have been specifically described in the AI context.

Data relevance is another mentioned property that can be subject to antipatterns.
Selected examples are the \textit{emphasis on available data}~\citeP{matthews2020patterns} or the usage of \textit{overcurated data}~\citeP{roselli2019managing}, both of which can lead to \textit{training/serving skew}.

One of the larger subcategories focuses on data management.
Since it can be quite complex in some cases, an \textit{undocumented data collection process}~\citeP{gudivada2017data} may negatively affect the long-term evolution of an AI-based system.
Similarly, \textit{premature data aggregation}~\citeP{munappy2019data} during this collection process can destroy important data points that cannot be retrieved again. 
A third example, which is frequently mentioned, is \textit{unsound/missing metadata}~\citeP{polyzotis2019data, munappy2019data, gudivada2017data}, i.e. a suboptimal or absent documentation and schema to describe the used~data.

Lastly, a smaller subcategory is related to data dependencies.
An example here is \textit{unstable data dependencies}~\citeP{Sculley2015primary, liu2020primary, belani2019requirements}.
Consuming data from other systems as input signals for model features may initially speed up development.
However, if the external data is unstable and changes over time, these dependencies can have negative and hard to diagnose effects on the related ML component.

\textit{Design \& Architecture Antipatterns.}
We also extracted eight antipatterns related to the design and architecture of AI-based systems.
While it is one of the smaller categories, it contains several frequently mentioned antipatterns.
An example related to modularity is \textit{undeclared consumers}~\citeP{Sculley2015primary, lwakatare2019taxonomy, belani2019requirements}, i.e. if the results of an AI component serve as input for a broad range of other systems or components.
These undeclared or silent consumers constitute hidden coupling, which can have negative and obscure side effects during software evolution.
Similarly, an antipattern where, instead of the output, the complete model is reused and slightly altered is called \textit{correction cascades}~\citeP{Sculley2015primary, matthews2020patterns, belani2019requirements}.
Changes in the original model then may lead to unintended ripple effects cascading to the \enquote{corrected} downstream models.
In~\citeP{matthews2020patterns}, this is also described with the improper reuse of complete AI components or pipelines.

A second type of antipatterns is concerned with software integration.
The prime example here is \textit{glue code}~\citeP{Sculley2015primary, alahdab2019empirical, lewis2019component}.
AI-based systems are often built with many generic packages or components, which are then connected with custom code, e.g. for data transformation or reading and writing data.
This not only makes it difficult to keep an overview of the system but glue code may also tightly couple the system to specific external libraries.
In the area of data collection and preparation, a specialized version of glue code is called \textit{pipeline jungles}~\citeP{Sculley2015primary, alahdab2019empirical}, i.e. the same stitching together but more on an architectural level and with ML pipelines.

Finally, a small subcategory regards technology adoption.
An example, not fully AI-specific, but still mentioned as a consequence of the nature of AI systems, is the \textit{multiple-language smell}~\citeP{Sculley2015primary, alahdab2019empirical, schelter2018challenges}. While using Python or R for ML models, and other languages for non-ML components, may enable using the best frameworks or libraries for the task at hand, it also entails disadvantages in maintaining, testing, or handing over a component to colleagues.

\textit{Code Antipatterns.}
We generally identified two subcategories of code antipatterns: those specific to AI-based systems and generic ones that occur more frequently in these systems.
A frequently mentioned AI-specific example is \textit{dead experimental codepaths}~\citeP{Sculley2015primary, alahdab2019empirical, arpteg2018software}.
The influence of data science leads to an iterative and experimental development process for AI components, where several conditional branches exist, which increase complexity and may also be forgotten, resulting in dead code.
Examples of generic code antipatterns which occur more frequently in AI software are \textit{long lambda functions} or \textit{long ternary conditional expressions}~\citeP{jebnoun2020scent}.

\textit{Infrastructure Antipatterns.}
We also identified a small number of antipatterns related to the infrastructure of AI-based systems.
The most prominent example from this category is \textit{weak or missing monitoring}~\citeP{breck2017primary, agarwal2016makingprimary, lewis2019component}.
Using AI components leads to additional observability requirements, e.g. monitoring data sources or model accuracy to detect \textit{training/serving skew}.
Moreover, the black-box nature of AI components can make it difficult to perform root cause analysis without specialized tooling.
Another infrastructure antipattern is hence \textit{weak or missing debugging}~\citeP{agarwal2016makingprimary}.
As a last example, \textit{inadequate configuration management}~\citeP{arpteg2018software} describes missing or suboptimal tooling mechanisms to manage important model configuration, e.g., features, preprocessing settings, or hyperparameters.

\textit{Ethics Antipatterns.}
The last smaller category of antipatterns is concerned with ethical issues in AI-based systems. 
The obvious example are \textit{biased models}~\citeP{breck2017primary, alahdab2019empirical}, i.e. models that have been created based on incomplete or irrelevant data or with a prejudice-inducing algorithm or process.
Such models not only produce inaccurate but also unfair results, which depending on the use case can have substantial negative societal effects, e.g. with predictive policing or recidivism models.
For such usage scenarios, it is especially important to measure and control the consequences of the respective AI system.
Failing to do so is described by the antipattern \textit{unmanaged social impact}~\citeP{matthews2020patterns}.
A final example in this category is \textit{undefined human accountability}~\citeP{matthews2020patterns}, i.e. when the role and responsibility of humans in AI-supported decisions is not clearly documented, allowing people to hide behind a~machine.

\begin{tcolorbox}
\textbf{Main findings (RQ2):} We extracted 72 unique antipatterns in six categories. Largest categories are \textit{model} (29) and \textit{data} (22). \textit{Design \& architecture} only consists of eight antipatterns, but many of them occurred several times. The antipattern which was mentioned the most is \textit{training/serving skew} (6).
\end{tcolorbox}

\subsection{Solutions (RQ3)}
From the 21 primary studies, we identified 46 unique instances of solutions.
Out of these, about a third of the instances explicitly mentions a TD type, another third mentions an antipattern, while the remaining third does not mention any specific TD type or antipattern addressed. 
In particular, the last group contains advices of broad and generic nature (e.g. \textit{perform extensive testing}~\citeP{breck2017primary}) and specific methods typically implemented in a tool or framework (e.g. \textit{Data Quality Advisor}~\citeP{shrivastava2019dqa}).
Given the difficulty in categorizing solutions using a single dimension (TD type, antipattern, or specificity level), we focus instead on discussing the five solutions that are most referenced in our primary studies (Table~\ref{table:solutions}).

\begin{table}
    \centering
	\caption{Most referenced solutions (occurred in at least 2 sources)}
	\label{table:solutions}
	\begin{tabular}{lrl}
		Solution Name & \# of Sources & Sources\\
		\hline
		\hline
		Manage model configuration & 3 & \citeP{schelter2018challenges, arpteg2018software, jebnoun2020scent}\\
		Use clear component and code APIs & 3 & \citeP{Sculley2015primary, alahdab2019empirical, o2020common}\\
        Remove unnecessary features & 2 & \citeP{Sculley2015primary, alahdab2019empirical}\\
        Refactor the code & 2 & \citeP{liu2020primary, jebnoun2020scent}\\
        Monitor deployed models & 2 & \citeP{schelter2018challenges, agarwal2016makingprimary}\\
		\hline
	\end{tabular}
\end{table}

\textit{Manage model configuration} prescribes that configuration changes in AI applications should be tracked, reviewed, and possibly tested in the same way as code~\citeP{arpteg2018software}. 
In this line, a suggested good practice is to externalize the configuration options from the code and to maintain them in human- and machine-readable files~\citeP{jebnoun2020scent}.
\textit{Use clear component and code APIs} relates instead to reducing design and architectural debt by encapsulating AI functionality in software components with clear required and provided interfaces~\citeP{alahdab2019empirical, o2020common}.
At the code level, wrapping black-box packages into custom APIs can address the \textit{glue code} antipattern~\citeP{Sculley2015primary}.
\textit{Remove unnecessary features} prescribes to reduce model debt by periodically examining if all input features of a model are still needed~\citeP{Sculley2015primary}, e.g. with so-called \enquote{leave-one-feature-out evaluations}. 
Additionally, new features should not be introduced if they do not significantly contribute to the prediction performance~\citeP{alahdab2019empirical}. 
\textit{Refactor the code} is a straightforward and generic solution to deal with design and code debt, e.g. in the form of code-related antipatterns~\citeP{jebnoun2020scent}. 
While this is a general best practice, AI-based systems require collaboration with expert developers when refactoring is needed, e.g. to boost performance or re-implement complex ML algorithms~\citeP{liu2020primary}.
Finally, \textit{monitor deployed models} suggests that ML models and their prediction performance should be closely monitored after deployment, e.g. to identify and address \textit{training/serving skew}~\citeP{agarwal2016makingprimary, schelter2018challenges}.

\begin{tcolorbox}
\textbf{Main findings (RQ3):} We extracted 46 unique solutions.
The solutions either explicitly address a TD type or an antipattern, or present a general best-practice to resolve TD in AI-based systems. 
The most referenced solutions are \textit{manage model configuration} (3) and \textit{use clear component and code APIs} (3).
\end{tcolorbox}

\section{Threats to Validity}
\label{sec:threats}
Several limitations have to be mentioned for our study.
\textit{Internal validity} is influenced by the applied scientific rigor and potentially hidden confounding factors, both of which may impact the consistency and correctness of the results.
Since selection, extraction, and synthesis activities of an SMS may rely partially on subjective interpretation, they may be prone to researcher bias.
Although we diligently designed and adhered to our SMS protocol and always assigned at least two researchers to each paper, other researchers may have achieved slightly different results with our protocol.

\textit{External validity} is concerned with the generalizability of the results.
With 21 final papers, our SMS can be regarded as comparatively small, which indicates that research on this topic is just getting started.
Moreover, many of our primary studies directly reference Sculley et al.~\citeP{Sculley2015primary} and build on their findings.
The majority of publications from industry is also from large software enterprises like Google, Amazon, or Microsoft.
Several results of our study are therefore heavily skewed towards Internet-scale ML systems.
As a consequence, reported facets of TD types or the relevance of certain antipatterns may slightly differ in other AI contexts.
 
\section{Conclusion}
\label{sec:conclusion}

In this paper, we aimed at characterizing the notions of TD and antipatterns for AI-based systems by performing a systematic mapping study. 
Our research questions focused on both established and new types of TD in these systems, but also on reported antipatterns and solutions. 
We identified four new TD types emerging in AI-based systems (data, model, configuration, and ethics debt) and observed that several established types (e.g. infrastructure, architectural, code, and test debt) are frequently occurring, although their scope was extended to include AI-specific aspects, e.g., the management and monitoring of both AI pipelines and models to mitigate infrastructure debt.
We also identified and categorized 72 unique antipatterns, the majority of which relate to data and models. 
Finally, we identified 46 solutions that can be used to reduce or prevent debt accumulation in~AI-based~systems.

For industry, our results can support AI/ML professionals to better communicate aspects of TD present in their systems, to raise awareness for common antipatterns, and to identify solutions to address both.
From a research perspective, our contribution provides an encompassing overview and characterization of TD and antipatterns that can emerge in the development of AI-based systems.
This study may also serve as a foundation for future research that both deepens our understanding of particular AI debt types and proposes more elaborate solutions to address them.
In this respect, we see potential for follow-up grey literature or interview studies in this area, as well as the development of tools and techniques to identify, address, or avoid specific AI antipatterns.

\balance
\bibliographystyle{ieeetr}
\bibliography{references}

\begin{thebibliography}{10}

\bibitem{agarwal2016makingprimary}
A.~Agarwal, S.~Bird, M.~Cozowicz, L.~Hoang, J.~Langford, S.~Lee, J.~Li,
  D.~Melamed, G.~Oshri, O.~Ribas, {\em et~al.}, ``Making contextual decisions
  with low technical debt,'' 2016.

\bibitem{lewis2019component}
G.~A. Lewis, S.~Bellomo, and A.~Galyardt, ``{Component Mismatches Are a
  Critical Bottleneck to Fielding AI-Enabled Systems in the Public Sector},''
  in {\em AAAI Fall Symposium Series: Artificial Intelligence in Government and
  Public Sector}, 2019.

\bibitem{Sculley2015primary}
D.~Sculley, G.~Holt, D.~Golovin, E.~Davydov, T.~Phillips, D.~Ebner,
  V.~Chaudhary, M.~Young, J.-F. Crespo, and D.~Dennison, ``{Hidden Technical
  Debt in Machine Learning Systems},'' in {\em International Conference on
  Neural Information Processing Systems}, 2015.

\bibitem{breck2017primary}
E.~Breck, S.~Cai, E.~Nielsen, M.~Salib, and D.~Sculley, ``{The ML Test Score: A
  Rubric for ML Production Readiness and Technical Debt Reduction},'' in {\em
  International Conference on Big Data}, IEEE, 2017.

\bibitem{o2020common}
K.~O'Leary and M.~Uchida, ``Common problems with creating machine learning
  pipelines from existing code,'' in {\em Machine Learning and Systems}, 2020.

\bibitem{polyzotis2019data}
N.~Polyzotis, M.~Zinkevich, S.~Roy, E.~Breck, and S.~Whang, ``Data validation
  for machine learning,'' vol.~1, 2019.

\bibitem{hynes2017data}
N.~Hynes, D.~Sculley, and M.~Terry, ``The data linter: Lightweight, automated
  sanity checking for ml data sets,'' in {\em MLSys Workshop}, 2017.

\bibitem{schelter2018challenges}
S.~Schelter, F.~Biessmann, T.~Januschowski, D.~Salinas, S.~Seufert, and
  G.~Szarvas, ``On challenges in machine learning model management,'' 2018.

\bibitem{shrivastava2019dqa}
S.~Shrivastava, D.~Patel, A.~Bhamidipaty, W.~M. Gifford, S.~A. Siegel, V.~S.
  Ganapavarapu, and J.~R. Kalagnanam, ``Dqa: Scalable, automated and
  interactive data quality advisor,'' in {\em International Conference on Big
  Data}, IEEE, 2019.

\bibitem{liu2020primary}
J.~Liu, Q.~Huang, X.~Xia, E.~Shihab, D.~Lo, and S.~Li, ``{Is Using Deep
  Learning Frameworks Free? Characterizing Technical Debt in Deep Learning
  Frameworks},'' in {\em International Conference on Software Engineering:
  Software Engineering in Society}, ACM/IEEE, 2020.

\bibitem{belani2019requirements}
H.~Belani, M.~Vukovic, and {\v{Z}}.~Car, ``Requirements engineering challenges
  in building ai-based complex systems,'' in {\em International Requirements
  Engineering Conference Workshops}, IEEE, 2019.

\bibitem{arpteg2018software}
A.~Arpteg, B.~Brinne, L.~Crnkovic-Friis, and J.~Bosch, ``Software engineering
  challenges of deep learning,'' in {\em Euromicro Conference on Software
  Engineering and Advanced Applications}, IEEE, 2018.

\bibitem{jebnoun2020scent}
H.~Jebnoun, H.~Ben~Braiek, M.~M. Rahman, and F.~Khomh, ``The scent of deep
  learning code: An empirical study,'' in {\em International Conference on
  Mining Software Repositories}, ACM, 2020.

\bibitem{matthews2020patterns}
J.~Matthews, ``Patterns and anti-patterns, principles and pitfalls:
  Accountability and transparency in ai,'' {\em AI Magazine, Association for
  the Advancement of Artificial Intelligence}, 2020.

\bibitem{munappy2019data}
A.~Munappy, J.~Bosch, H.~H. Olsson, A.~Arpteg, and B.~Brinne, ``{Data
  Management Challenges for Deep Learning},'' in {\em Euromicro Conference on
  Software Engineering and Advanced Applications}, IEEE, 2019.

\bibitem{gudivada2017data}
V.~Gudivada, A.~Apon, and J.~Ding, ``Data quality considerations for big data
  and machine learning: Going beyond data cleaning and transformations,''
  vol.~10, 2017.

\bibitem{alahdab2019empirical}
M.~Alahdab and G.~{\c{C}}al{\i}kl{\i}, ``Empirical analysis of hidden technical
  debt patterns in machine learning software,'' in {\em International
  Conference on Product-Focused Software Process Improvement}, Springer, 2019.

\bibitem{vakkuri2020just}
V.~Vakkuri, K.-K. Kemell, M.~Jantunen, and P.~Abrahamsson, ``{“This is Just a
  Prototype”: How Ethics Are Ignored in Software Startup-Like
  Environments},'' in {\em International Conference on Agile Software
  Development}, Springer, 2020.

\bibitem{roselli2019managing}
D.~Roselli, J.~Matthews, and N.~Talagala, ``Managing bias in {AI},'' in {\em
  Companion Proceedings of the World Wide Web Conference}, 2019.

\bibitem{lwakatare2019taxonomy}
L.~E. Lwakatare, A.~Raj, J.~Bosch, H.~H. Olsson, and I.~Crnkovic, ``{A taxonomy
  of software engineering challenges for machine learning systems: An empirical
  investigation},'' in {\em International Conference on Agile Software
  Development}, Springer, 2019.

\bibitem{foidl2019risk}
H.~Foidl and M.~Felderer, ``Risk-based data validation in machine
  learning-based software systems,'' in {\em {International Workshop on Machine
  Learning Techniques for Software Quality Evaluation}}, 2019.

\end{thebibliography}


\begin{thebibliography}{10}

\bibitem{francois_chollet_deep_2017}
{François Chollet}, {\em Deep {Learning} with {Python}}.
\newblock Manning, 2017.

\bibitem{amershi_software_2019}
S.~Amershi, A.~Begel, C.~Bird, R.~DeLine, H.~Gall, E.~Kamar, N.~Nagappan,
  B.~Nushi, and T.~Zimmermann, ``Software {Engineering} for {Machine}
  {Learning}: {A} {Case} {Study},'' in {\em {International} {Conference} on
  {Software} {Engineering}: {Software} {Engineering} in {Practice}},
  {IEEE}/{ACM}, 2019.

\bibitem{rahman_machine_2019}
M.~S. Rahman, E.~Rivera, F.~Khomh, Y.-G. Guéhéneuc, and B.~Lehnert, ``Machine
  {Learning} {Software} {Engineering} in {Practice}: {An} {Industrial} {Case}
  {Study},'' {\em arXiv:1906.07154 [cs]}, June 2019.
\newblock arXiv: 1906.07154.

\bibitem{Sculley2015}
D.~Sculley, G.~Holt, D.~Golovin, E.~Davydov, T.~Phillips, D.~Ebner,
  V.~Chaudhary, M.~Young, J.-F. Crespo, and D.~Dennison, ``{Hidden Technical
  Debt in Machine Learning Systems},'' in {\em International Conference on
  Neural Information Processing Systems}, 2015.

\bibitem{bosch_engineering_2021}
J.~Bosch, I.~Crnkovic, and H.~H. Olsson, {\em Engineering {AI} {Systems}: {A}
  {Research} {Agenda}}, pp.~1--19.
\newblock {IGI Global}, 2021.

\bibitem{RN2274}
W.~Cunningham, ``The {WyCash} {Portfolio} {Management} {System},'' in {\em
  OOPSLA proceedings}, 1992.

\bibitem{avgeriou2016managing}
P.~Avgeriou, P.~Kruchten, I.~Ozkaya, and C.~Seaman, ``{Managing technical debt
  in software engineering (Dagstuhl Seminar 16162)},'' in {\em Dagstuhl
  Reports}, vol.~6, Schloss Dagstuhl-Leibniz-Zentrum fuer Informatik, 2016.

\bibitem{agarwal2016making}
A.~Agarwal, S.~Bird, M.~Cozowicz, L.~Hoang, J.~Langford, S.~Lee, J.~Li,
  D.~Melamed, G.~Oshri, O.~Ribas, {\em et~al.}, ``Making contextual decisions
  with low technical debt,'' 2016.

\bibitem{breck2017ml}
E.~Breck, S.~Cai, E.~Nielsen, M.~Salib, and D.~Sculley, ``{The ML Test Score: A
  Rubric for ML Production Readiness and Technical Debt Reduction},'' in {\em
  International Conference on Big Data}, IEEE, 2017.

\bibitem{liu2020using}
J.~Liu, Q.~Huang, X.~Xia, E.~Shihab, D.~Lo, and S.~Li, ``{Is Using Deep
  Learning Frameworks Free? Characterizing Technical Debt in Deep Learning
  Frameworks},'' in {\em International Conference on Software Engineering:
  Software Engineering in Society}, ACM/IEEE, 2020.

\bibitem{fairbanks2020ur}
G.~Fairbanks, ``Ur-technical debt,'' {\em IEEE Annals of the History of
  Computing}, vol.~37, no.~04, 2020.

\bibitem{li2015systematic}
Z.~Li, P.~Avgeriou, and P.~Liang, ``A systematic mapping study on technical
  debt and its management,'' Elsevier, 2015.

\bibitem{verdecchia2020architectural}
R.~Verdecchia, P.~Kruchten, P.~Lago, and I.~Malavolta, ``Building and
  evaluating a theory of architectural technical debt in software-intensive
  systems,'' Journal of Systems and Software. Elsevier, 2021.

\bibitem{guo2011portfolio}
Y.~Guo and C.~Seaman, ``A portfolio approach to technical debt management,'' in
  {\em Workshop on Managing Technical Debt}, ACM, 2011.

\bibitem{tamburri2013social}
D.~A. Tamburri, P.~Kruchten, P.~Lago, and H.~van Vliet, ``What is social debt
  in software engineering?,'' in {\em International Workshop on Cooperative and
  Human Aspects of Software Engineering}, IEEE, 2013.

\bibitem{rios2018tertiary}
N.~Rios, M.~G. de~Mendon{\c{c}}a~Neto, and R.~O. Sp{\'\i}nola, ``A tertiary
  study on technical debt: Types, management strategies, research trends, and
  base information for practitioners,'' vol.~102, Elsevier, 2018.

\bibitem{Brown1998}
W.~Brown, R.~Malveau, H.~S. McCormick, and T.~Mowbray, {\em {AntiPatterns:
  Refactoring Software, Architectures, and Projects in Crisis}}.
\newblock New York, NY, USA: John Wiley {\&} Sons, Inc., 1st~ed., 1998.

\bibitem{Fowler1999}
M.~Fowler, {\em {Refactoring: Improving the Design of Existing Code}}.
\newblock Addison-Wesley Professional, 1999.

\bibitem{Garcia2009}
J.~Garcia, D.~Popescu, G.~Edwards, and N.~Medvidovic, ``{Identifying
  Architectural Bad Smells},'' in {\em 2009 13th European Conference on
  Software Maintenance and Reengineering}, IEEE, 2009.

\bibitem{Khomh2009}
F.~Khomh, M.~{Di Penta}, and Y.-G. Gueheneuc, ``{An Exploratory Study of the
  Impact of Code Smells on Software Change-proneness},'' in {\em Working
  Conference on Reverse Engineering}, IEEE, 2009.

\bibitem{Fontana2016}
F.~A. Fontana, J.~Dietrich, B.~Walter, A.~Yamashita, and M.~Zanoni,
  ``{Antipattern and Code Smell False Positives: Preliminary Conceptualization
  and Classification},'' in {\em International Conference on Software Analysis,
  Evolution, and Reengineering}, IEEE, mar 2016.

\bibitem{Bogner2019-CSEQUDOS}
J.~Bogner, T.~Boceck, M.~Popp, D.~Tschechlov, S.~Wagner, and A.~Zimmermann,
  ``{Towards a Collaborative Repository for the Documentation of Service-Based
  Antipatterns and Bad Smells},'' in {\em International Conference on Software
  Architecture Companion}, IEEE, 2019.

\bibitem{Albarak2018}
M.~Albarak and R.~Bahsoon, ``{Prioritizing technical debt in database
  normalization using portfolio theory and data quality metrics},'' in {\em
  International Conference on Technical Debt}, ACM Press, 2018.

\bibitem{Foidl2019a}
H.~Foidl, M.~Felderer, and S.~Biffl, ``{Technical Debt in Data-Intensive
  Software Systems},'' in {\em Euromicro Conference on Software Engineering and
  Advanced Applications}, IEEE, 2019.

\bibitem{Humbatova2020}
N.~Humbatova, G.~Jahangirova, G.~Bavota, V.~Riccio, A.~Stocco, and P.~Tonella,
  ``{Taxonomy of real faults in deep learning systems},'' in {\em International
  Conference on Software Engineering}, ACM, 2020.

\bibitem{Felderer2021}
M.~Felderer and R.~Ramler, ``{Quality Assurance for AI-Based Systems: Overview
  and Challenges},'' in {\em Software Quality: Future Perspectives on Software
  Engineering Quality. SWQD 2021. Lecture Notes in Business Information
  Processing}, vol.~404, Springer International Publishing, 2021.

\bibitem{Santhanam2020}
P.~Santhanam, ``{Quality Management of Machine Learning Systems},'' in {\em
  Communications in Computer and Information Science}, vol.~1272, Springer
  International Publishing, 2020.

\bibitem{Serban2020}
A.~Serban, K.~van~der Blom, H.~Hoos, and J.~Visser, ``{Adoption and Effects of
  Software Engineering Best Practices in Machine Learning},'' in {\em
  International Symposium on Empirical Software Engineering and Measurement},
  ACM, oct 2020.

\bibitem{Siebert2020}
J.~Siebert, L.~Joeckel, J.~Heidrich, K.~Nakamichi, K.~Ohashi, I.~Namba,
  R.~Yamamoto, and M.~Aoyama, ``{Towards Guidelines for Assessing Qualities of
  Machine Learning Systems},'' in {\em Quality of Information and
  Communications Technology}, Springer, 2020.

\bibitem{Washizaki2019}
H.~Washizaki, H.~Uchida, F.~Khomh, and Y.-G. Gueheneuc, ``{Studying Software
  Engineering Patterns for Designing Machine Learning Systems},'' in {\em
  International Workshop on Empirical Software Engineering in Practice}, IEEE,
  dec 2019.

\bibitem{Kitchenham2004}
B.~Kitchenham, ``{Procedures for performing systematic reviews},'' {\em Keele,
  UK, Keele University}, vol.~33, no.~TR/SE-0401, 2004.

\bibitem{Wohlin2014}
C.~Wohlin, ``{Guidelines for snowballing in systematic literature studies and a
  replication in software engineering},'' in {\em International Conference on
  Evaluation and Assessment in Software Engineering}, ACM Press, 2014.

\bibitem{Petersen2008}
K.~Petersen, R.~Feldt, S.~Mujtaba, and M.~Mattsson, ``{Systematic mapping
  studies in software engineering},'' 2008.

\bibitem{ISO25010}
{International Organization For Standardization}, ``{ISO/IEC 25010 - Systems
  and software engineering - Systems and software Quality Requirements and
  Evaluation (SQuaRE) - System and software quality models},'' 2011.

\bibitem{jenner2004companion}
B.~Jenner, U.~Flick, E.~von Kardoff, and I.~Steinke, {\em A companion to
  qualitative research}.
\newblock Sage, 2004.

\end{thebibliography}
\vspace{5pt}

\bibliographystyleP{ieeetr}
\bibliographyP{references}


\end{document}